# Ultrasonic elastography for the prevention of breast implant rupture: detection of an increase with stiffness over implantation time

Laetitia Ruffenach[a,b], Dimitri Heintz[c], Claire Villette[c], Charlène Cosentino[a], Denis Funfschilling[a], Frédéric Bodin[a,b], Nadia Bahlouli[a], Simon Chatelin[a,*]

[a] *ICube, UMR 7357 CNRS, University of Strasbourg, Strasbourg, France*
[b] *Service de chirurgie plastique esthétique et reconstructrice, Hautepierre hospital, CHRU Strasbourg, HUS, Strasbourg, France*
[c] *Plant Imaging & Mass Spectrometry (PIMS), IBMP, UPR 2357 CNRS, University of Strasbourg, Strasbourg, France*

**Abstract**

Breast implants are widely used after breast cancer resection and must be changed regularly to avoid a rupture. To date, there are no quantitative criteria to help this decision. The mechanical evolution of the gels and membranes of the implants is still under investigated, although it can lead to early rupture. In this study, 35 breast explants having been implanted in patients for up to 17 years were characterized by ex vivo measurements of their mechanical properties. Using Acoustic Radiation Force Impulse (ARFI) ultrasound elastography, an imaging method for non-destructive mechanical characterization, an increase in the stiffness of the explants has been observed. This increase was correlated with the implantation duration, primarily after 8 years of implantation. With an increase of the shear modulus of up to a factor of nearly 3, the loss of flexibility of the implants is likely to lead to a significant increase of their risk of rupture. A complementary analysis of the gel from the explants by mass spectrometry imaging (MSI) and liquid chromatography coupled to high resolution mass spectrometry (LC-HRMS) confirms the presence of metabolites of cholesterol originating from the breast tissues, which most likely crossed the membrane of the implants and most likely degrades the gel. By observing the consequences of the physical-chemical mechanisms at work within patients, this study shows that ultrasound elastography could be used in vivo as a quantitative indicator of the risk of breast implant rupture and help diagnose their replacement.

*Keywords:* breast implants, Acoustic Radiation Force Impulse, ultrasound elasticity imaging

*Corresponding authors at: the engineering science, computer science and imaging laboratory (ICube), UMR 7357 CNRS, University of Strasbourg, 1 place de l'hôpital, 67000 Strasbourg, France.
E-mail addresses: schatelin@unistra.fr (S. Chatelin)

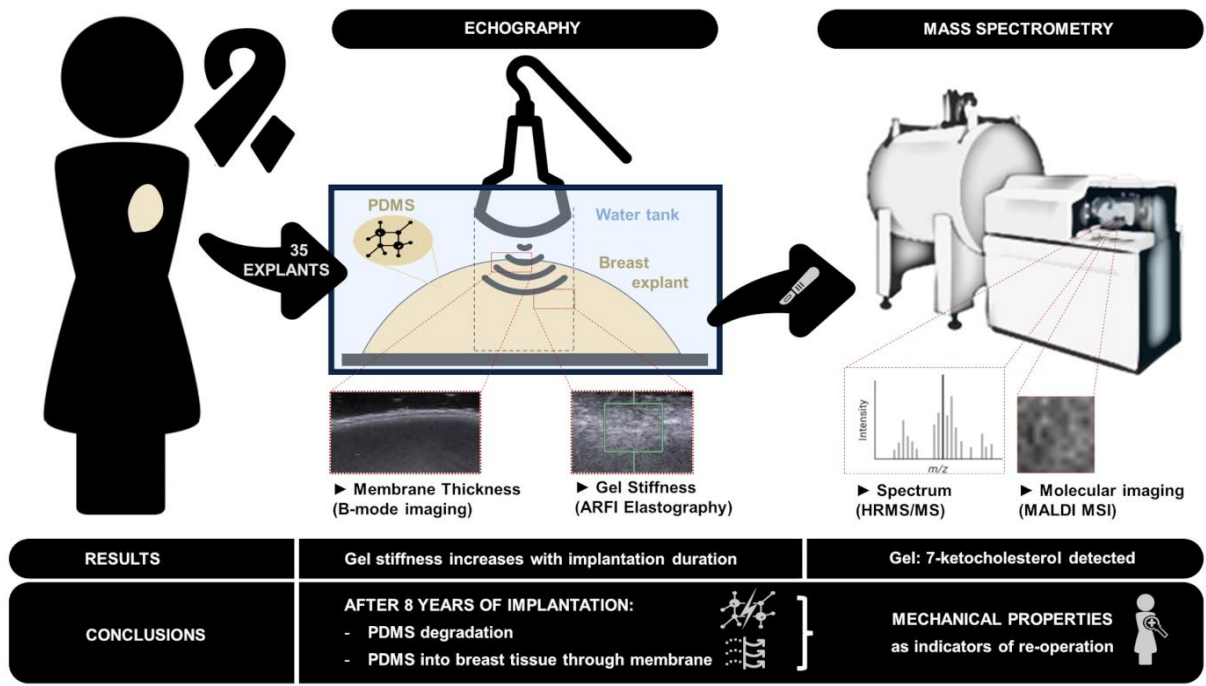

## 1. Introduction

Of the more than 58,000 people affected by breast cancer yearly in France, total surgical breast removal (mastectomy) occurs in one-third of the cases (*Institut Curie*, France) with 30% of patients undergoing breast reconstruction post-mastectomy (*Haute Autorité de Santé*, 2020) as part of their cancer treatment (*American Cancer Society*, 2019). On average, reconstruction requires reoperation every 10 years and possible links between implants and complications/mortality have been identified (Abi-Rafeh et al., 2019). As recommended by the American Food and Drug Administration, it is therefore essential to study the mechanisms of implant aging (Le-Petross et al., 2023) in order to propose new predictive diagnostic tools to limit risk of rupture and re-operation.

Nearly 70,000 implants are sold each year in France (*ANSM*, France), and are primarily composed of silicone shells (PolyDiMethylSiloxane (PDMS) and a layer of diphenyl/dimethylsilicone or fluorosilicone (Necchi et al., 2011)) containing a gel (lightly crosslinked silicone elastomer and ~80% of high molecular mass PDMS (Brook, 2006)). While breast implant shells must meet the NF EN ISO 14607-14630 standards, the standard is not indicative of the mechanical load on the implant and thus cannot be used to predict implant lifespan. Previous studies correlate implant rupture and implantation duration (Bodin et al., 2015; Brandon et al., 2002), due to micro-changes and stiffening of the membrane (Magill et al., 2019).

The ability for the membrane to prevent exchange between the gel and surrounding tissue is critical for both patient and the implant as bleeding, the migration of low molecular mass silicone oil, has a weakening effect of membrane mechanical strength. Although 85% of breast implants between 2007 and 2016 were textured implants (*ANSM*, France), surgeons now prefer to use smooth implants to reduce the risk of anaplastic large cell lymphoma (Rastogi et al., 2018). Another major issue with membrane permeability is the possibility of lipid infiltration from surrounding biological tissue into the implants (Adams et al., 1998), although there is currently no information on how this affects mechanical characteristics.

Over the past 30 years, elastography has emerged as an effective non-invasive diagnostic tool (Bercoff et al., 2003; Lewa, 1991; Muthupillai et al., 1995; Sandrin et al., 2002; Sarvazyan et al., 1998), especially for the detection of breast cancer. Elastography has been used to assess the impact of breast implants on the biological environment such as inflammation (Rzymski et al., 2011), capsular contracture (Kuehlmann et al., 2016; Prantl et al., 2014; Sowa et al., 2017), fibrosis (Jung et al., 2021) or hematoma (de Faria Castro Fleury et al., 2017). A first study suggested elastography for implant surveillance, showing mechanical differences between intact pre-operative and ruptured implants (Stachs et al., 2015).

Our objective was to analyze to use of elastography for the evaluation of non-ruptured breast implants and analyzed the evolution of mechanical properties over implantation time. We hypothesized that the stability of PDMS is degraded by components from the surrounding tissues, leading to changes in the mechanical properties, an increased risk of rupture, and a potential release of synthetic components into the body.

**2. Methods**

*2.1. Description of the samples*

In this study, we recovered explanted breast implants (explants) from patients who had undergone post-mastectomy reconstruction surgery a few years earlier. 26 explants were collected by the *Aesthetic and Reconstructive Plastic Surgery Department* of Strasbourg Hospital, following the current regulations and legislation. The mechanical properties of 35 silicone prothesis explants were measured in this study, 9 of them were new and 26 were previously implanted and then explanted in patients. Data concerning the history of the prosthesis were collected. Following explantation, the explants were stored individually in hermetically sealed boxes at room temperature for a period ranging from 3-33 months, which we assume is unlikely to have altered the properties of the explants. Only unbroken specimens have been extracted and tested. The explants were divided into four groups, according to their manufacturer: 24 textured "*Allergan* 410" specimens (*Allergan Inc., Irvine, CA, USA*) (called *Allergan* samples), 3 smooth "*Cerecare* Ellipse RVH" specimens (*Cerecare®, Sailly-lez-Cambrai, France*) (called *Cerecare* samples), 6 smooth "*Sebbin* LSC" specimens (*Laboratoires Sebbin, Boissy-l'Aillerie, France*) (called *Sebbin* samples) and 2 smooth "*Sebbin Sizer*" explants usually used for preoperative placement (called *Sebbin Sizer*). The data related to respective implantation time were collected (**Table 1**). Some patients underwent simultaneous bilateral explantation of two explants (12 of the explants collected, specifically marked as shaded cells in Table 1). Of the explants, implantation duration ranged from 24-210 months.

**Table 1.** Summary, referencing and history of the different implants used in this study. Shaded cells indicate two explants coming from the same patient.

| Sample # [-] | Manufacturer | Mass [g] | Implantation time [months] | Patient # [-] | Delay between explantation and test [months] |
|---|---|---|---|---|---|
| 1 | *Allergan Inc.* | Unknown | 121 | 1 | 9 |
| 2 | *Allergan Inc.* | Unknown | 121 | 1 | 9 |
| 3 | *Allergan Inc.* | Unknown | 132 | 2 | 2 |
| 4 | *Allergan Inc.* | Unknown | 99 | 3 | 7 |
| 5 | *Allergan Inc.* | Unknown | 88 | 4 | 6 |
| 6 | *Allergan Inc.* | Unknown | 202 | 5 | 9 |
| 7 | *Allergan Inc.* | Unknown | 76 | 6 | 6 |
| 8 | *Allergan Inc.* | Unknown | 76 | 6 | 6 |
| 9 | *Allergan Inc.* | Unknown | 121 | 7 | 5 |
| 12 | *Allergan Inc.* | 410 | 128 | 8 | 13 |
| 15 | *Allergan Inc.* | 410 | 122 | 9 | 7 |
| 16 | *Allergan Inc.* | 285 | 181 | 10 | 1 |
| 17 | *Allergan Inc.* | 315 | 115 | 11 | 6 |
| 18 | *Allergan Inc.* | 315 | 115 | 11 | 6 |
| 19 | *Allergan Inc.* | 375 | 177 | 12 | 19 |
| 20 | *Allergan Inc.* | 370 | 38 | 13 | 33 |
| 21 | *Allergan Inc.* | 220 | 38 | 13 | 33 |
| 22 | *Allergan Inc.* | 230 | 141 | 14 | 11 |
| 23 | *Allergan Inc.* | 275 | 120 | 15 | 18 |
| 24 | *Allergan Inc.* | 195 | 120 | 16 | 13 |
| 25 | *Allergan Inc.* | 430 | 210 | 17 | 12 |
| 26 | *Allergan Inc.* | 255 | 122 | **18** | 18 |
| 27 | *Allergan Inc.* | 255 | 122 | **18** | 18 |
| 28 | *Allergan Inc.* | 255 | 140 | 19 | 23 |
| 29 | *Laboratoires Sebbin* | 330 | 0 (New) | - (New) | - (New) |
| 30 | *Laboratoires Sebbin* | 385 | 24 | **20** | 3 |
| 31 | *Laboratoires Sebbin* | 385 | 24 | **20** | 3 |
| 32 | *Laboratoires Sebbin* | 240 | 0 (New) | - (New) | - (New) |
| 33 | *Laboratoires Sebbin* | 475 | 0 (New) | - (New) | - (New) |
| 34 | *Laboratoires Sebbin* | 265 | 0 (New) | - (New) | - (New) |
| 35 | *Cerecare®* | 280 | 0 (New) | - (New) | - (New) |
| 36 | *Cerecare®* | 250 | 0 (New) | - (New) | - (New) |
| 37 | *Cerecare®* | 340 | 0 (New) | - (New) | - (New) |
| 38 | *Laboratoires Sebbin* | 280 | 0 (New) (*Sizer*) | - (New) | - (New) |
| 39 | *Laboratoires Sebbin* | 245 | 0 (New) (*Sizer*) | - (New) | - (New) |

*2.2. Ultrasonography protocol*

The explants were immersed in water to facilitate ultrasound transmission, as illustrated in velocities $c_s$ are measured and then expressed in terms of shear modulus µ assuming linear elastic waves with $\mu = \rho c_s^2$. The density ρ is assumed to be 1,030 kg.m$^{-3}$ for PDMS at room temperature, which is a common value found in literature and manufacturer's datasheet (Genovés et al., 2023). For each explant, silicone gel stiffness values are given as averaged shear modulus from 3 distinct regions of interest (ROIs) and 10 values for each ROI at depths ranging from 10 to 15 mm below the membrane.

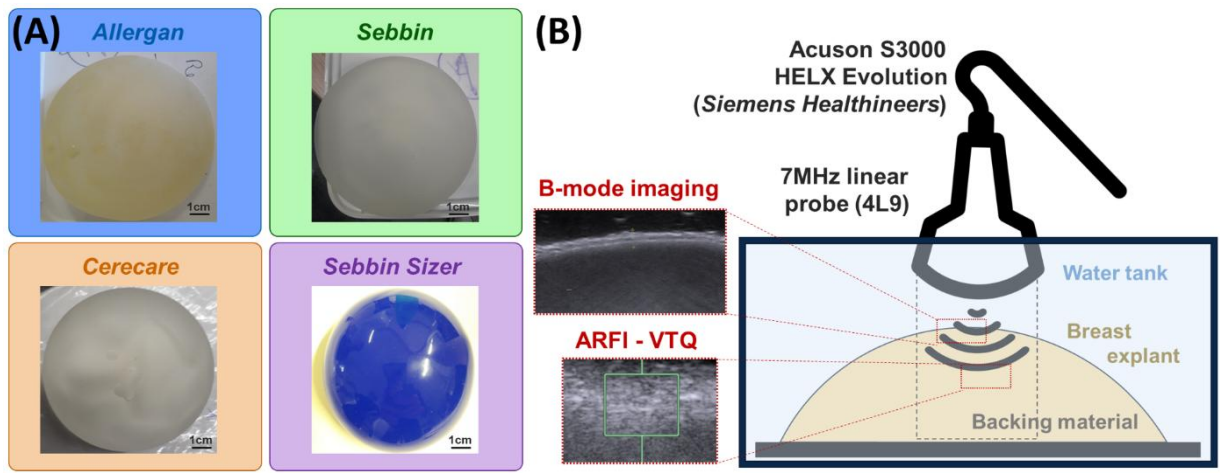

Fig. *1*. The samples were scanned using a clinical ultrasound system (*Acuson S3000 HELX Evolution, Siemens Healthineers, Erlangen, Germany*) with a 4L9 linear array transducer centered at 7 MHz. The multi-layer membrane of each explant was observed, and their thicknesses measured using B-mode imaging. Based on the generation of shear waves by focused ultrasound in the organ and then the detection of their propagation velocity by ultrasound imaging, ARFI (Acoustic Radiation Force Impulse) imaging methods (Nightingale et al., 2002, 2001; Palmeri et al., 2008) allows to extract non-invasively the mechanical properties. The gel stiffness was then quantified using Virtual Touch Quantification (VTQ), a shear wave elastography method based on ARFI imaging. Shear wave

*2.3. Statistical analysis*

Both membrane thickness and gel stiffness values are expressed as Mean ± Standard Deviation (SD). For comparison between groups of explants by manufacturer and in view of the small number of samples in some groups, we used non-parametric Kruskal–Wallis (one-way ANOVA on ranks) tests for samples to show any significant differences between two groups. For all the statistical analyses, a p-value of less than 0.05 was considered to be significant. The correlation between the implant stiffness and the implantation duration has been evaluated using the Pearson (PCC) and the Spearman's ($r_S$) Rank Correlation Coefficients. The aim is to estimate a potential linear correlation and to assess a monotonic link, respectively, between these two parameters on the 24 *Allergan* samples.

*2.4. MALDI-MSI protocol*

After the ultrasound tests, two *Allergan* explants (#6 and #7) were used for matrix assisted laser desorption/ionization mass spectrometry imaging (MALDI-MSI) analysis (Fig. 2). The aim is to determine: (1) whether organic compounds from the surrounding biological breast tissues were likely to be found in the gel of the explants after having crossed the membrane; (2) whether they would be able to provide hypotheses as to the modification of the mechanical properties of the inner gel.

The gels were deposited on an indium tin oxide (ITO) glass slide. As membranes would not stick to the glass slide, a conductive copper tape was used to make imprints of the membrane's surfaces. Then, the samples were covered with α-cyano-4-hydroxycinnamic acid (HCCA) matrix (10 mg/mL in 70% acetonitrile, 0.2% trifluoroacetic acid) using a M5 spotter (HTX). MALDI-MSI acquisitions were performed in positive ion mode on a 100-1,000 Da mass range using a MALDI-FT-ICR mass spectrometer (SolariX XR 7.0T, *Bruker Daltonics GmbH, Bremen, Germany*) set at 2M. The laser power was set to 20% with 200 laser shots and a frequency of 1,000 Hz, with a raster size of 50x50 µm. The instrument was calibrated before the analysis on the peaks of HCCA matrix, and on-line calibration was used to calibrate each acquired spectra during the acquisition.

Images were displayed in flexImaging 5.0. The spectra were annotated in MetaboScape® 2022b (*Bruker Daltonics GmbH, Bremen, Germany*) with a maximum mass deviation of 3 ppm and a maximum mSigma value of 30, assessing the good fitting of isotopic profiles. The proposed annotations were checked by liquid chromatography coupled to high resolution mass spectrometry (LC-HRMS/MS) operating in positive ion mode on a 100 to 1,000 Da mass range, as described in (Silvestro et al., 2018). The gels and membranes were extracted using 600 µL of methanol/chloroform (2/1) on 0.6 g of sample, vortexed and incubated 1h before injection. The information obtained from the fragmentation of the compounds of interest in AutoMS/MS mode allowed the interrogation of spectral libraries. Annotations are proposed to the level 2 of Schymanski classification (Schymanski et al., 2014). The quantification of 7-ketocholesterol was performed using the analytical standard of the compound obtained from Avanti Polar Lipids. The standard was prepared at 1, 2.5, 5 and 7.5 µg/mL in chloroform and deposited on an ITO glass slide. The spots were covered with HCCA matrix and analyzed by MALDI-MSI using the same methods as for the samples. A standard curve was established in SCiLS Lab Pro software with a $R^2$=0.98.

## 3. Results

*3.1. Characterization of the membrane*

The different layers of the membrane appeared on B-mode imaging with 2 layers for *Allergan* and *Cerecare* and 3 to 5 layers for *Sebbin* (

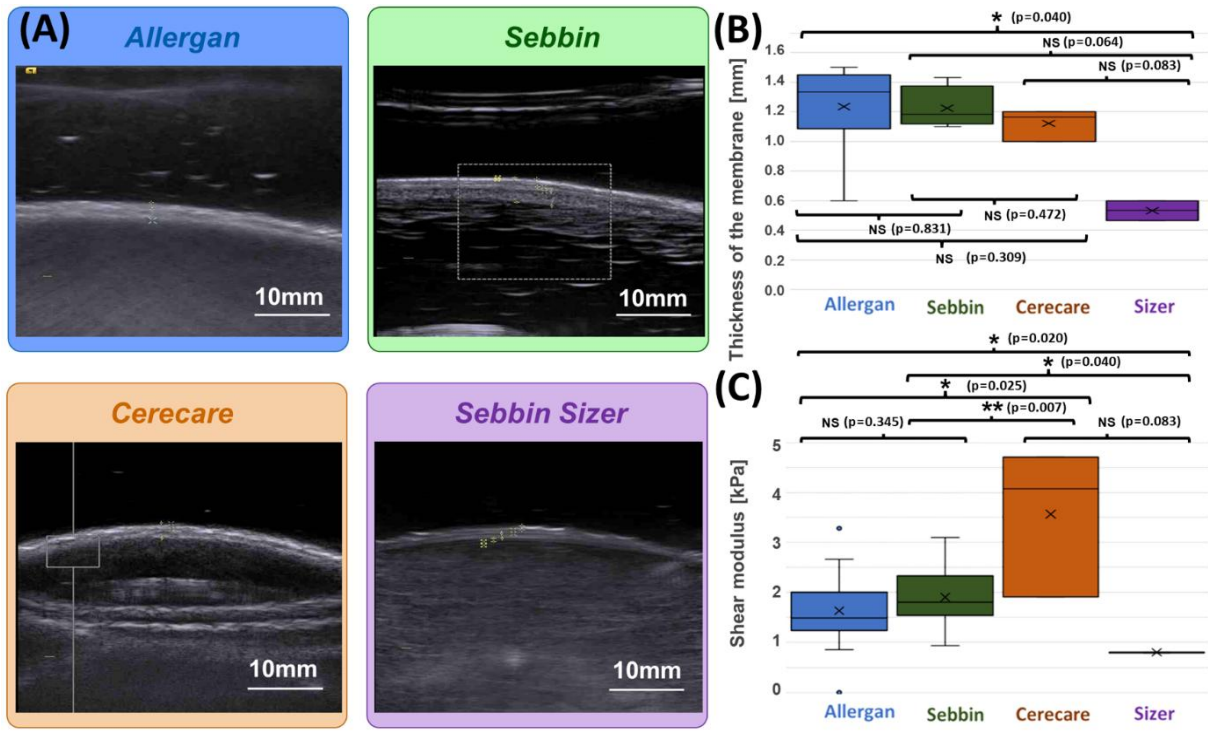

Fig. **3**(A)). The membranes had a thickness of 1.2 ± 0.2 mm, 1.2 ± 0.1 mm, 1.1 ± 0.1 mm and 0.5 ± 0.1 mm for *Allergan*, *Sebbin*, *Cerecare* and *Sebbin Sizer*, respectively (

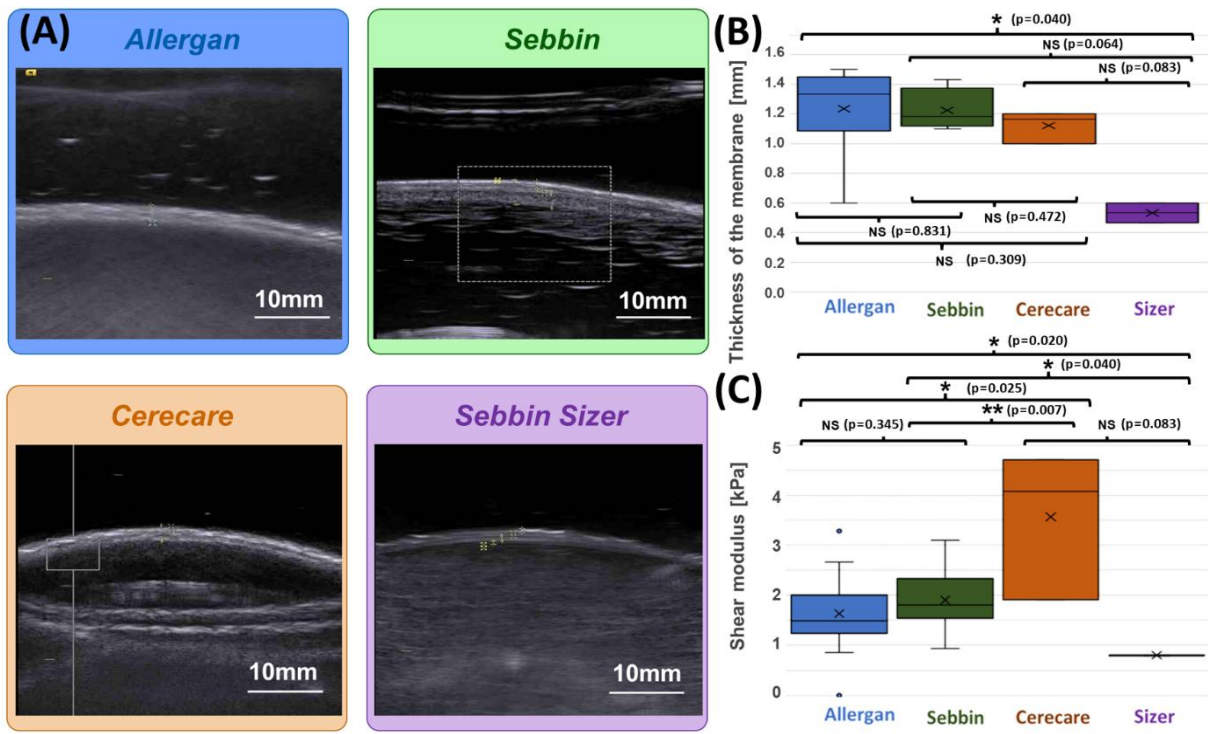

Fig. 3(B)). While the thicknesses were similar for the three categories of explants intended for long-term implantation, the thickness of the *Sebbin Sizer* for short-term implantation was lower.

*3.2. Inter-manufacturer variability of the gel stiffness*

The measurements of the stiffness of the filling gel shows significant differences from one brand to another (

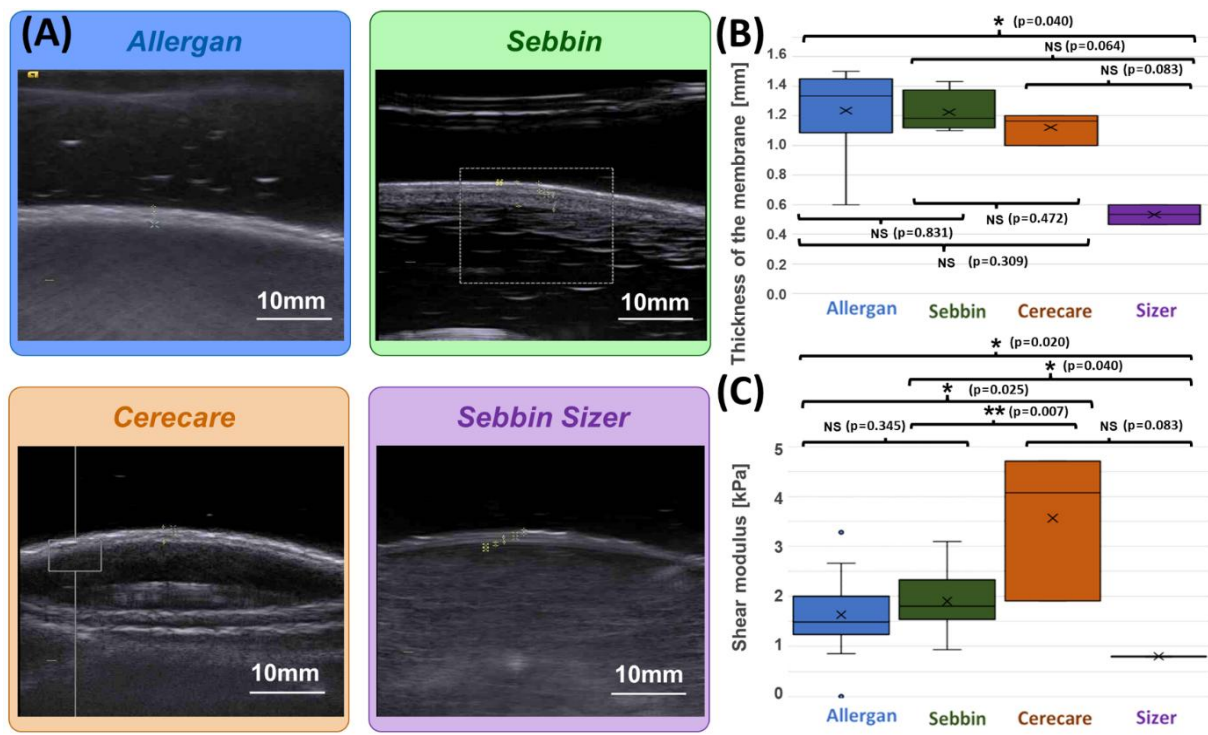

Fig. 3(C)). Thus, while *Allergan* and *Sebbin* have similar stiffness (1.79 ± 0.71 kPa and 1.90 ± 0.67 kPa, respectively), *Cerecare* implants appear stiffer (3.57 ± 1.47 kPa). The *Sebbin Sizer* implants have the lower stiffness (0.80 ± 0.01 kPa). We can assume that the manufacturing processes of the *Sebbin* and *Sebbin Sizer* have an influence on the gel stiffness, as the latter are not intended for long-term resistance. Unlike the *Sebbin Sizer*, the values on *Cerecare* show large standard deviation.

*3.3. Variations over implantation duration*

For *Allergan*, we had 5 pairs of explants, extracted from the same patient two by two. A pair-wise comparison of the stiffness of these explants shows similar values (Fig. 4(A)). This is not the case for the only pair of *Sebbin*. For *Sebbin* a real significant difference appears between the two implants. Fig. 4(B) shows in blue the *ex vivo* analysis of the stiffness of the silicone gel of the 24 *Allergan* explants (textured) over their implantation duration.

The mechanical properties of all the explants tested showed a significant increase of stiffness in time after about 8 years of implantation and this increase is correlated with the implantation duration (PCC = 0.67, $r_S$ = 0.5). An explant carried for 210 months is 3 times stiffer than for 38 months. This quasi-linear behavior in two parts has an inflection point around 8 years of implantation (PCC of 0.72 and 0.85 for <7 and >9 years, respectively). The evolution of the shear modulus is compared to that of the rupture risk (Kaplan and Meier, 1958) reported in the literature (Caplin et al., 2021; Maxwell et al., 2015; Spear and Murphy, 2014). Although only a few prospective systematic analyses have been conducted (Bodin et al., 2015), a similar trend is observed.

First differences appeared by visual observation of the general external aspect of the explants, the oldest ones often presenting an accentuated yellowish aspect, suggesting exchanges with the biological tissue (Fig. 4(B)).

*3.4. Mass spectroscopy imaging (MSI) analysis*

From MALDI-MSI, Phosphoglycerides were found on the membranes and metabolites of cholesterol (3-ketocholesterol and 7-ketocholesterol - 3KC and 7KC, respectively) originating from the breast tissues were detected homogeneously within the gel (Fig. 2). This was confirmed in LC-HRMS/MS, showing that these organic molecules were able to cross the membranes and diffuse in the gel. The amount of 7KC was quantified at 2.36 ± 0.63 µg/mL and 0.92 ± 0.38 µg/mL in samples #6 and #7, respectively.

**4. Discussions**

We have evaluated non-destructively the mechanical properties of the gel for many explants of different brands and with known implantation history. The explants tested in this study do not concern those that have been ruptured, as those cannot be recovered within the framework of research protocols for reasons of administrative, insurance or legal issues. We described the evolution of the implants with a sufficiently high number of measurements to attest the significance of the results. We assume that storage at room temperature in individual hermetically sealed boxes is unlikely to further alter the properties of the samples tested.

The thickness of the membranes remains similar between the different brands, despite potential differences in the manufacturing process. No changes in membrane thickness were observed as a function of implantation duration. Despite the strong interest that this could present for the prediction of the risk of rupture, it is not currently possible to apply the usual elastography methods to characterize the mechanical properties of the membranes. Because of a too thin structure, methods used routinely in clinical exams are still less performant and could lead to greatly distorted values. This is explained by the low thickness of the membrane which generates guided wave phenomena during the application of these methods by dynamic elastography.

Concerning the stiffness of the inner gel, although it is obtained by reticulation of two silicone polymer components coming *a priori* from the same manufacturers, significant differences appear, which suggests a potential influence in the manufacturing process of the implants. This is confirmed by the fact that the gel contained in the *Sebbin Sizer* is softer than the gel coming in *Sebbin* (same manufacturer), but which are intended for long-term implantation. In addition to properties related to the prosthesis manufacturing process itself, intrinsic biological factors linked to the treatments and medical and pathological history undergone by the patient during implantation (such as disease or medication) could eventually play a role in certain disparities in measured properties.

The most striking result is the temporal evolution of the stiffness, with a significant and very clear stiffening after about 8 years of implantation. This observation and timelapse seem to be correlated with the fact that rupture rates, initially very low, begin to increase after 6 to 8 years of implantation (Hillard et al., 2017). A similar stiffening was observed for the prosthesis from the same patient. In addition to the risk of wear over time of the membrane, this stiffening of the silicone gel is likely to lead to a lack of mechanical flexibility of the implant over time and thus increase the risk of rupture.

While the membranes are supposed to guarantee the impermeability of the prostheses, MALDI-MSI of two of the explants clearly detected the presence of cholesterol metabolites 3KC and 7KC, originating from the breast tissues, in large quantities and in a relatively homogeneous manner. 7KC is a toxic oxysterol that disrupts lipid packing, causes cholesterol depletion and can be assumed to have a direct effect on PDMS stiffness as it has been shown for biological tissues such as endothelial cells or vascular walls (Levitan et al., 2014). Although potentially not the only mechanism at work, the presence of 7KC could thus explain the increase in stiffness observed by elastography. However, at this stage we have not yet been able to extend these results to all explants, nor to compare them in similar explants that have not yet been worn. Their quantification, shown in a preliminary way in this manuscript, is one of the major perspectives of this preliminary study and would require sampling at different sites for each explant in a precise, repeatable and comparable way.

There are two main limitations to these results concerning the influence of the implantation duration. Both come from the fact that prostheses that are currently implanted are not textured any more like those evaluated over time. First, the possible exchanges through the membranes and the stiffening of the gel would be potentially different between textured and non-textured implants. Then, we have no more access to new and unworn textured explants. This did not allow us to set up a control group and thus to fully conclude whether the changes in stiffness were related to the ageing of the implants (due to the solicitations and exchanges in the patients' bodies) or independently of these and occurring even outside the body. A similar study on non-textured implants is a major prospect for extending the present study, with a control group not worn and followed over time. However, even if the textured prostheses are no longer implanted, they are still in place in many patients today and the question of a criterion for estimating their explantation is still widely discussed.

The prostheses were tested *ex vivo*, after explantation. Even immerged in water, the ultrasound signal was relatively low, and it was not possible for most explants to measure stiffness at depths more than 1 to 2 cm below the membrane. We can reasonably assume that this is due both to a significant attenuation of ultrasound in the silicone gel, but also to the presence of a low ultrasound speckle in this gel, as well as to the presence of the membrane reflecting part of the signal. This last point is confirmed by a higher Signal to Noice Ratio (SNR), especially in *Sebbin Sizer* implants which have a much thinner membrane. It constitutes a limiting point of the use of *in vivo* elastography for the noninvasive estimation of the stiffness of breast implants. This limitation should be reconsidered because we observed a higher SNR in the case of non-textured explants which are implanted very often now and for which measurements are possible at a much larger depth.

This preliminary study opens the way for an understanding of the mechanical phenomena leading to the rupture of the implants, but also to the perspective of the use of non-invasive elastography to predict this risk of rupture and, therefore, help to diagnose the need for implant replacement.

**Acknowledgements**

This work was funded by the *Interdisciplinary Thematic Institute HealthTech*, as part of the ITI 2021-2028 program of the University of Strasbourg, CNRS and Inserm, was also supported by the IHU Strasbourg (*Institute of Image Guided Surgery, ANR-10-IAHU-02*), by IdEx Unistra (ANR-10-IDEX-0002) with the IRIS technology platform and by SFRI (STRAT'US project, ANR-20-SFRI-0012) under the framework of the *French Investments for the Future Program*. In figure 2, breast implant and scalpel, mass spectrometer and mass spectrum were created with BioRender.com. The author gratefully acknowledges Nicole Kirsch for her help in proofreading the manuscript.

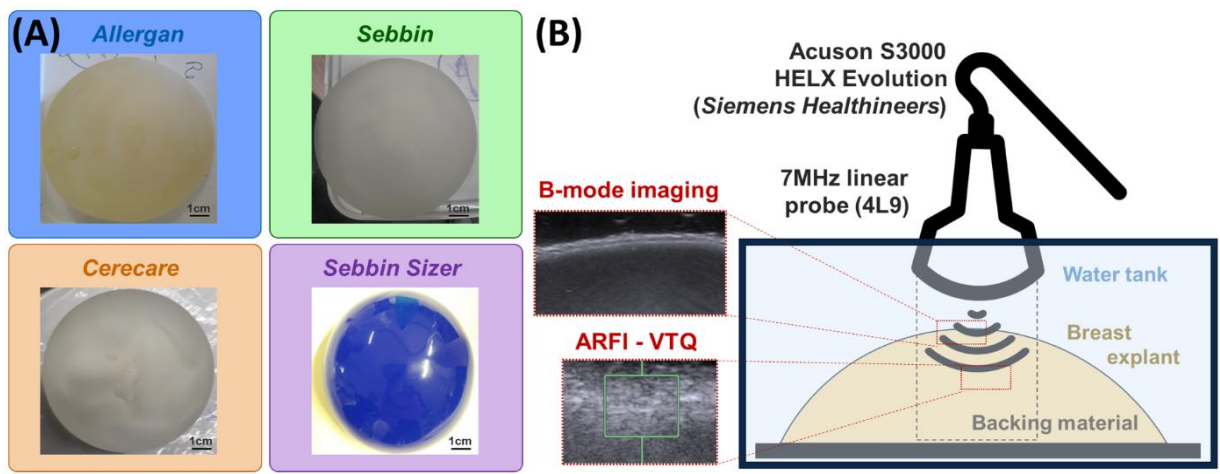

**Fig. 1.** (A) Example of explants from the 4 manufacturers. (B) Illustration of the experimental configuration. The explants were immersed in water. The membrane thickness was measured using B-mode imaging and the gel stiffness quantified by ARFI VTQ elastography.

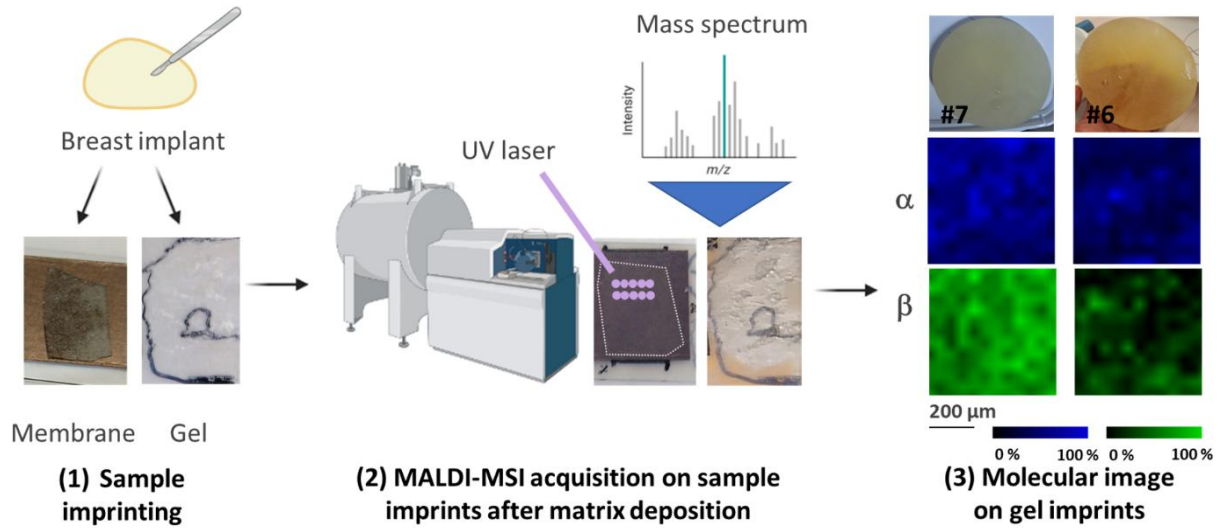

**Fig. 2.** MALDI-MSI workflow for the molecular analysis of membrane and gel from breast implants. (1) Membrane and gel were separated and directly deposited on an ITO-coated glass slide (gel) or imprinted on a copper tape covering the slide (membrane). (2) MALDI ionization matrix was applied before shooting the samples with a laser to obtain molecular images from membrane and gel. Molecules were identified by data base interrogation. (3) MALDI-MSI performed on the gel from *Allergan* textured samples #6 and #7 indicate presence of 3-ketocholesterol (α) and 7-ketocholesterol (β).

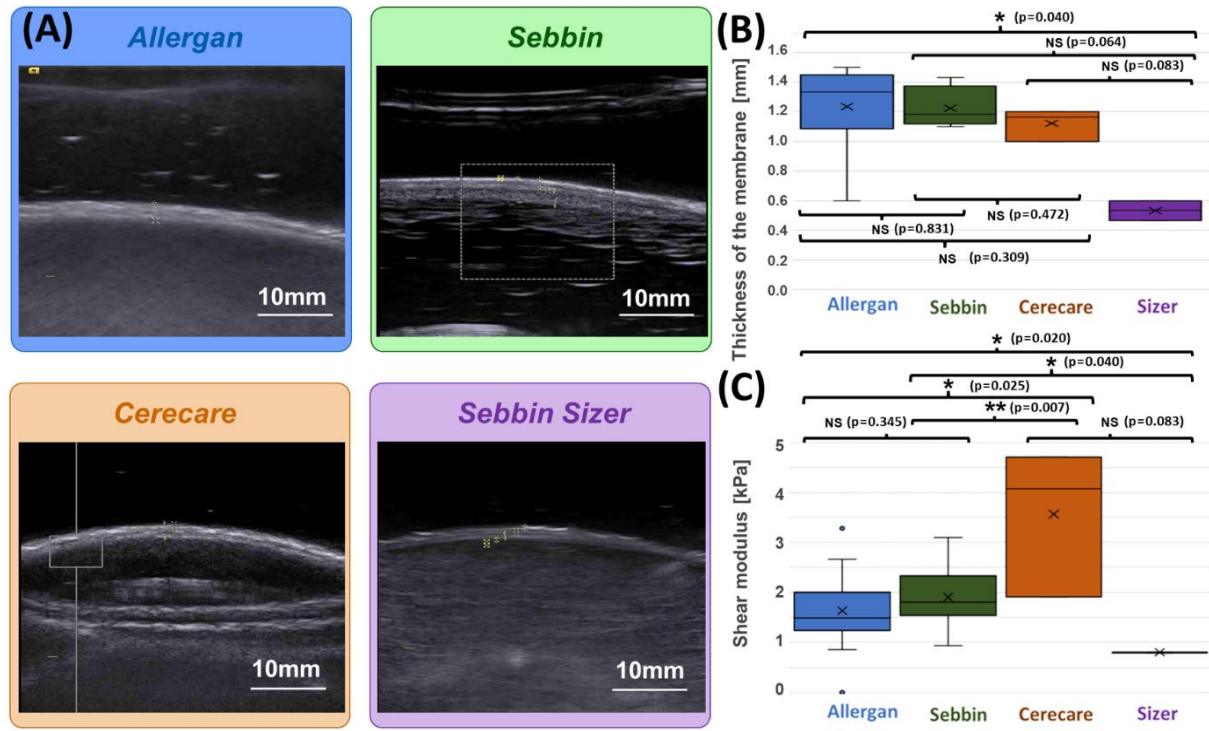

**Fig. 3.** (A) Examples of B-mode images for the measurement of membrane thickness of implants from each of the four types of implants. mages are presented at the same scale. Membrane thicknesses (B) and then shear moduli measured by ARFI method (C) are compared without dissociation as a function of prosthesis age or implantation time. (*NS: p-value > 0.05, \*: p-value < 0.05, \*\*: p-value < 0.01, \*\*\*: p-value < 0.001*)

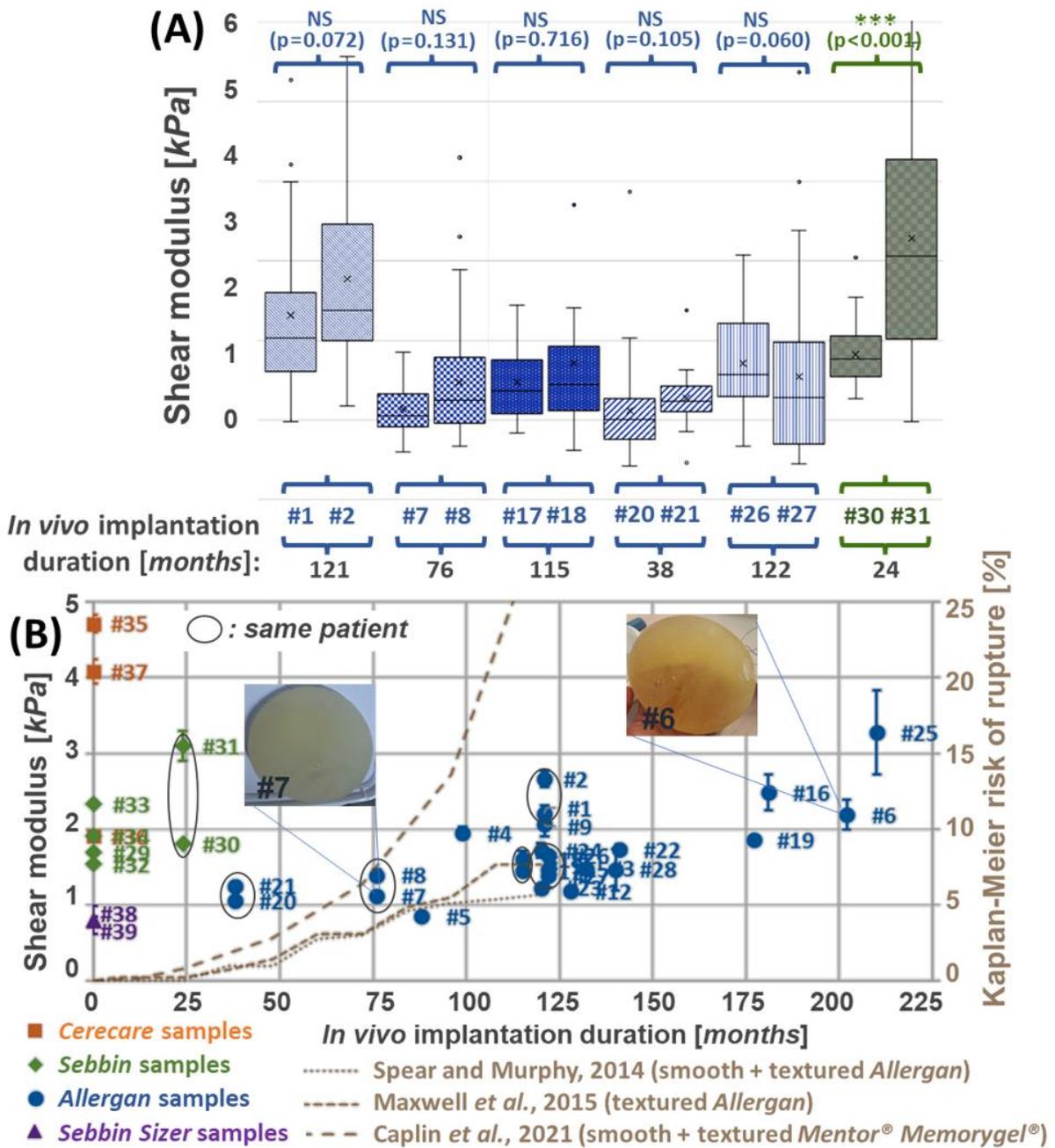

**Fig. 4.** (A) Paired comparison of stiffness values by ARFI measurements of explants from the same patients. Explant shear modulus is presented as a function of implantation duration for *Allergan* textured in blue and *Sebbin* / *Cerecare* / *Sebbin Sizer* smooth implants in green, orange and purple, respectively, collected in this study (B) (primary axis). A clear correlation between implantation time and implant stiffness appeared significantly (PCC = 0.67, $r_s$ = 0.5). The evolution of the shear modulus is compared to that of the rupture risk (Kaplan-Meier risk rate (Kaplan and Meier, 1958)) reported in the literature (brown, secondary axis) (Caplin et al., 2021; Maxwell et al., 2015; Spear and Murphy, 2014). *(NS: p-value > 0.05, \*: p-value < 0.05, \*\*: p-value < 0.01, \*\*\*: p-value < 0.001)*

19